# Dynamics of Triaxial Galaxies

By TIM DE ZEEUW[1]

[1]Sterrewacht Leiden, Postbus 9513, 2300 RA Leiden, The Netherlands

Luminous galaxies and their dark halos are likely to have triaxial shapes. The construction of distribution functions for triaxial systems is a hard problem, due to the existence of only one exact integral of motion, the energy $E$. Even so, progress has been made by use of analytic methods for special potentials, numerical combination of orbit densities (Schwarzschild's method), and N-body simulations. Despite these efforts, we are a long way from understanding – let alone constructing – the full variety of models needed to interpret the photometric and kinematic data on galaxies that is now available. Some of the main unsolved problems include the range of intrinsic shapes and figure rotation speeds for which stable dynamical equilibria can be built, and the dynamical role of cusps and central black holes.

## 1. Introduction

Many galaxy components are not spherical or even axisymmetric, but are instead triaxial: cosmological N-body simulations routinely produce triaxial dark halos; giant elliptical galaxies have been known to be slowly-rotating triaxial structures for two decades; there is mounting evidence that many bulges, including the one in our own Galaxy, are triaxial; bars have long been known to be ferociously triaxial (e.g., Binney 1976; Kormendy 1982; Frenk et al. 1988; Aguilar 1988; Blitz & Spergel 1991; Barnes 1992).

Key questions for scenarios of galaxy formation are: (i) What is the distribution of intrinsic triaxial shapes? (ii) What is, at a given shape, the range in internal velocity distributions? (iii) Do galaxies and their halos occupy the whole set of shapes that is possible according to dynamics, or do they form only a subset of all possible stable equilibria? (iv) What is the role of dark halos? (v) What is the dynamical role of massive central black holes? (vi) What is the distribution of metals within galaxies, and what is the relation between the detailed kinematics of stars (and gas) with the local metal enrichment?

These questions can be answered by morphological, kinematical and stellar population studies of galaxies as functions of redshift, when combined with the predictions of detailed theoretical models. In particular, dynamical modeling provides information on the intrinsic shapes of galaxies, on the motions of the stars and gas, and hence on the connection between the kinematics and the physical properties of stars and interstellar medium. Dynamical modeling also establishes the importance of dark halos and of massive black holes in galaxy nuclei.

This paper reviews the current state of knowledge on the dynamics of triaxial systems, with attention to some of the key contributions by Donald Lynden–Bell. Implications for the structure of elliptical galaxies are discussed in some detail, and the next steps in this field are outlined.

## 2. Dynamics

An equilibrium dynamical model of a galaxy is fully specified by its phase-space distribution function (DF) $f(\mathbf{x}, \mathbf{v}) \geq 0$, which gives the density of stars at each position $\mathbf{x}$ and velocity $\mathbf{v}$. Jeans' theorem states that $f$ depends on the six phase-space coordinates $\mathbf{x}$ and $\mathbf{v}$ through the isolating integrals of motion admitted by the gravitational potential of the system (Lynden–Bell 1962b). Since there are generally at most three such independent integrals, this reduces the number of variables in the problem of finding $f$. The isolating integrals of motion label the individual stellar orbits. Jeans' theorem therefore states that a dynamical model is defined once the number of stars that populate each orbit is specified.

The DF of a galaxy must be reconstructed from its observed properties. Since we cannot resolve the individual stars in all but the nearest galaxies, the only available observables are those integrated along the line of sight (at each position on the sky). The volume density has to be recovered from the surface brightness $\Sigma$, while the internal kinematics must be deduced from the line-of-sight velocity distribution, or velocity profile, $VP(v_{\rm los})$. Measurements of VPs require high signal-to-noise ratio spectroscopic data. Until recently only the lowest order moments of the VP, namely the mean line-of-sight velocity $\langle v_{\rm los}\rangle$, and the velocity dispersion $\sigma_{\rm los}$, could be recovered from the observations. These can be derived from the lowest order velocity moments of the DF, $\langle v_i^k v_j^\ell\rangle$, defined as integrals of $v_i^k v_j^\ell f$ over all velocities (with $i,j = x, y, z$, and $k, \ell = 0, 1, 2$). The mean streaming motions $\langle v_j\rangle$ (which project to $\langle v_{\rm los}\rangle$) satisfy a continuity equation; the second moments $\langle v_i v_j\rangle$ (connected to $\sigma_{\rm los}$) are related to the density and the potential of the model by the Jeans equations (e.g., Binney & Tremaine 1987). Solutions of the Jeans equations have provided some insight into the dynamics of elliptical galaxies. However, these studies have a major limitation: the Jeans solutions are not guaranteed to be physical, i.e., to correspond to a DF that is everywhere non-negative. Knowledge of the entire DF is required to derive physical dynamical conclusions. In addition, the entire DF is needed for the computation of theoretical VPs. These are essential for the interpretation of the VP measurements that have become feasible through recent technological progress (see de Zeeuw 1994 for a recent review).

Application of Jeans' theorem requires knowledge of the isolating integrals of motion. This is why it is most useful for spherical galaxies, in which the energy $E$ and the angular momentum $\mathbf{L}$ are conserved. Most spherical models that have been constructed have DFs of the form $f(E, L^2)$, so that the velocity distribution is anisotropic, but does not have a preferred tangential direction (Dejonghe 1986; Richstone & Tremaine 1988; Gerhard 1991). Including a dependence of the DF on $L_z$ allows mean streaming around the $z$-axis (e.g., Lynden–Bell 1960a). Jeans' theorem is also useful for axisymmetric models with $f = f(E, L_z)$, as first shown by Lynden–Bell (1962a). The actual calculation of such DFs has been beset by technical difficulties (Dejonghe 1986), which have recently been solved (e.g., Hunter & Qian 1993). Examples of realistic $f(E, L_z)$-models for axisymmetric galaxies can be found in, e.g., Dehnen & Gerhard (1994) and Qian et al. (1995).

## 3. Triaxial systems

Triaxial potentials generally admit only one exact isolating integral, the orbital energy $E$. Although there are three planes of reflection symmetry, there are no symmetry axes, and no component of the angular momentum vector is conserved. Thus, two of the three integrals of motion are unknown, and may not exist for all orbits (see §3.4). This prohibits a direct application of Jeans' theorem, and is a serious obstacle in the construction of equilibrium models, i.e., in the construction of physical DFs. Two approaches have been followed to gain insight in the dynamics of triaxial systems:

1. Numerical construction of individual models, either by N-body simulations (Aarseth & Binney 1976; Wilkinson & James 1982), or by Schwarzschild's (1979) method. In the latter approach individual orbits are calculated numerically in a chosen potential, and are then populated so as to reproduce the associated density. Both methods were important in establishing the existence of equilibrium models with centrally concentrated density distributions, and stationary or slowly tumbling triaxial shapes (see the reviews by de Zeeuw & Franx 1991; Barnes & Hernquist 1992; Bertin & Stiavelli 1993).

2. Semi-analytical study of special classes of models whose properties bracket the general case, and for which sufficient simplification occurs so that whole families of them can be studied at once. Three such families are known:

• The separable models, which have density profiles with cores, a variety of shapes, and stationary figures. All orbits in these models have three exact integrals of motion, and Jeans' theorem can be applied.

• The scale-free models, in which all properties vary as simple powers of the radius, but the only exact integral is the energy $E$. They are simple models for cusps.

• The power-law galaxies. These have potentials stratified on ellipsoids, allow cores and cusps, and have a generic orbital structure. Many of the observable properties can be computed easily.

### 3.1. *Separable models*

Lynden–Bell (1962c) provided a comprehensive list of non-axisymmetric potentials which admit three exact isolating integrals of motion $E$, $I_2$ and $I_3$. The most general are those for which the Hamilton–Jacobi equation separates in confocal ellipsoidal coordinates. The resulting integrals $I_2$ and $I_3$ depend quadratically on the velocity components, and are related to the angular momentum integrals of the axisymmetric and spherical limits (de Zeeuw & Lynden–Bell 1985). These *separable potentials* were first classified systematically by Stäckel (1890), and made their appearance in astronomy through Eddington's (1915) work. For a long time, attention was restricted to the axisymmetric limiting case, in which $I_2$ reduces to $L_z$, while the third integral $I_3$ remains non-classical (e.g., Kuzmin 1956; Ollongren 1962). The main application was in understanding the motion of stars in our own Galaxy, which was known to involve a third integral (Contopoulos 1960). No DFs were constructed, until very recently (Dejonghe & de Zeeuw 1988; Dejonghe 1992; Dejonghe et al. 1995).

Three fundamental properties of the general separable models make them useful for studies of galaxies:

($a$) They correspond to smooth inhomogeneous mass models, with non-rotating triaxial shapes, an arbitrary short-axis density profile, and arbitrary central axis ratios (Kuzmin 1973; de Zeeuw 1985c). Models in which the density profile falls off $\propto r^{-4}$ at large radii $r$ remain flattened as $r \to \infty$. When the density profile has a shallower slope, the density distribution becomes more nearly spherical as $r$ increases (de Zeeuw, Peletier & Franx 1986). On projection, the ellipticity of the isophotes generally changes with radius, but they show no twisting (Franx 1988). All relevant models have constant density cores.

($b$) The three-dimensional orbital motion is tractable analytically, and is a combination of three one-dimensional motions, each of which is either an *oscillation* or a *rotation* in one of the three ellipsoidal coordinates in which the motion separates. Different combinations of oscillations and rotations result in four families of orbits: boxes, short-axis tubes, inner long-axis tubes, and outer long-axis tubes (Kuzmin 1973; de Zeeuw 1985b). The box orbits have no net circulation and penetrate to the center. The tubes carry net angular momentum around either the short axis or the long axis. Illustrations of the four orbital shapes are given in Statler (1987).

($c$) The four orbit families are identical to the four major orbit families found in Schwarzschild's (1979) non-rotating triaxial model. They are generic for stationary triaxial models with a core, and moderate flattening (de Zeeuw 1985a; Gerhard 1985).

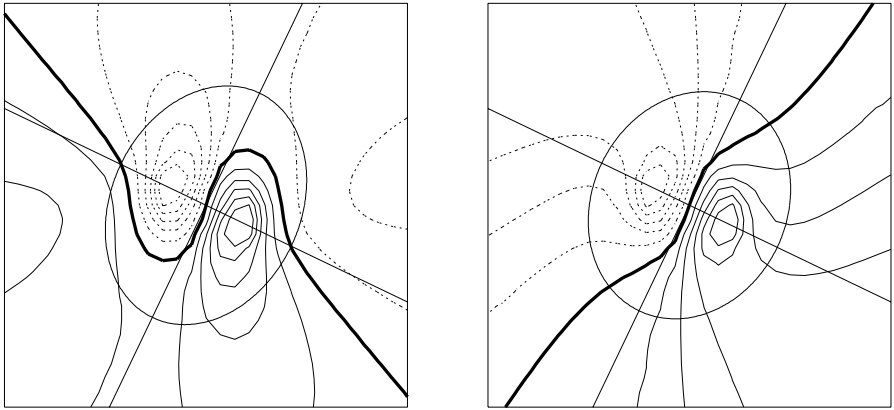

FIGURE 1. Examples of line-of-sight projected velocity fields $\langle v_{\rm los}\rangle$ of triaxial galaxies. The thin ellipse is a surface brightness isophote of the model. The thick curve indicates zero velocity. The dotted contours indicate negative velocities (approaching), and the solid contours indicate positive velocities (receding). Conventional long-slit spectroscopy is usually restricted to the major- and minor axis of the galaxy image, indicated by the thin straight lines (Arnold, de Zeeuw & Hunter 1994).

Lynden–Bell (1960b) was the first to write down the fundamental integral equation that connects a DF $f(E, I_2, I_3)$ to the separable triaxial density $\rho(x, y, z)$. He suggested that this equation might have a unique solution, and wrote: "While there is, as yet, no certainty that non-rotating stellar systems without axial symmetry exist, the struggle for a solution of the above equation [relating $f$ to $\rho$] does not seem worth the effort."

This situation persisted for more than 20 years. Kuzmin's (1973) four-page paper presented one set of smooth separable triaxial mass models, and established the existence of the four orbit families. However, it remained unknown in the West until 1985. It was in the summer of 1982 that Donald Lynden-Bell suggested to me that I take a look at the potentials that separate in confocal ellipsoidal coordinates. My subsequent visit to the famous office "01" at the IoA (Evans, this conference) then put me on the road to the rediscovery of the mass models, and their orbital structure. Comparison with the known properties of Schwarzschild's numerical models for elliptical galaxies (known to be triaxial since Binney's 1976 paper) quickly showed that the separable models might shed substantial light on the dynamics of these systems, so that it became interesting to find solutions of the fundamental integral equation.

Direct calculation of distribution functions $f(E, I_2, I_3)$ consistent with a given triaxial density $\rho(x, y, z)$ by solution of the fundamental integral equation is indeed difficult, as Lynden-Bell suspected, and requires a lot of effort (Dejonghe & Laurent 1991; Hunter & de Zeeuw 1992). However, contrary to Lynden–Bell's expectation, the solutions are highly non-unique. The individual orbital densities in a separable model are known explicitly, so that it is straightforward to build selfconsistent models by means of Schwarzschild's method. By using this approach, Statler (1987) found that the presence of four major orbit families provides ample opportunity for exchanging orbits of different shapes while keeping the same model density (but varying the observable kinematic properties).

In addition to the fundamental non-uniqueness that arises from the exchange of different orbits, there is a further non-uniqueness caused by the freedom to choose the direction of circulation for each star on each tube orbit. As a result, net streaming is possible around the short axis and around the long axis, and the observed line-of-sight mean streaming velocities $\langle v_{\rm los}\rangle$ display a rich kinematic structure (Statler 1991a, 1994a, b; Arnold, de Zeeuw & Hunter 1994). The zero velocity curve generally is misaligned from the apparent minor axis (Figure 1). This means that there is a velocity gradient along this axis. Such gradients ('minor-axis rotation') have been observed in many giant elliptical galaxies (Figure 2). Complete velocity reversals ('counter-rotation') can occur in certain radial intervals along some position angles for specific viewing angles, even when the stars on tube orbits all have the same sense of rotation (Statler 1991b). A number of elliptical galaxies display such reversals (e.g., Franx & Illingworth 1988; Jedrzejewski & Schechter 1988). In many cases these are probably not caused by projection, but are the signature of *intrinsic* kinematic decoupling: since the sense of rotation on any long- or short-axis tube is free, it is easy to construct stationary triaxial models with an inner part that rotates in the opposite sense to the outer part, or at right angles to it. Various scenarios have been proposed for the formation of such systems (see de Zeeuw & Franx 1991).

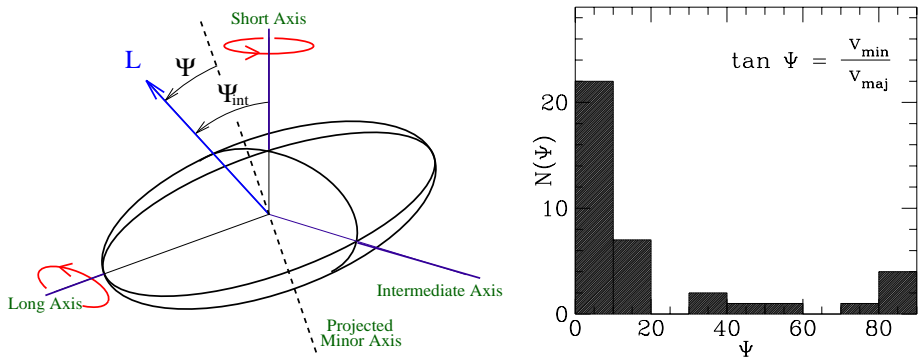

FIGURE 2. Minor axis rotation. Left panel: In a triaxial model the only orbits with net angular momentum circulate around either the long or the short axis. The total angular momentum vector **L** therefore lies in the plane containing these two axes, and is generally misaligned from the short axis by an angle $\Psi_{\rm int}$. When observed in projection, there is a velocity gradient along the apparent minor axis, since this is misaligned from the rotation axis by an angle $\Psi$. Right panel: Distribution of $\Psi$ for a sample of 38 elliptical galaxies (Franx, Illingworth & de Zeeuw 1991).

### 3.2. *Scale-free models*

The second set of special models that has received attention in the last decade are those with density profiles that are single power laws, so that the properties at one radius are related to those at another radius by a simple scaling. This simplifies the construction of selfconsistent equilibria. Most studies have considered oblate and stationary triaxial models with $\rho \propto r^{-2}$, which can be thought of as deformed singular isothermal spheres.

The orbital structure of non-tumbling scale-free triaxial models with $\rho \propto r^{-\gamma}$ differs fundamentally from that of models with homogeneous cores, which include all the separable systems. The three major families of tube orbits still exist, and hence the mean streaming motions are similar to those described in §3.1. However, some or all of the box orbits are replaced by minor orbit families and irregular orbits (Gerhard & Binney 1985; Udry & Pfenniger 1988). These are associated with stable and unstable higher-order resonances between the oscillation frequencies along and perpendicular to the principal axes, and have been christened *boxlets* by Miralda-Escudé & Schwarzschild (1989).

Boxlets display a large variety of shapes, but in substantially flattened triaxial models they do not provide enough density close to the long axis, as the very elongated box orbits that remain close to this axis do. These boxes are needed in any selfconsistent triaxial model, because all tube orbits are elongated opposite to the figure of the model, and hence cannot by themselves reproduce the density distribution. This suggests that whereas oblate selfconsistent scale-free models do exist (Richstone 1980, 1984; Toomre 1982; Levison & Richstone 1985; Evans 1993, 1994), flattened triaxial equilibria of this kind might not (Lees & Schwarzschild 1992), at least not for all shapes.

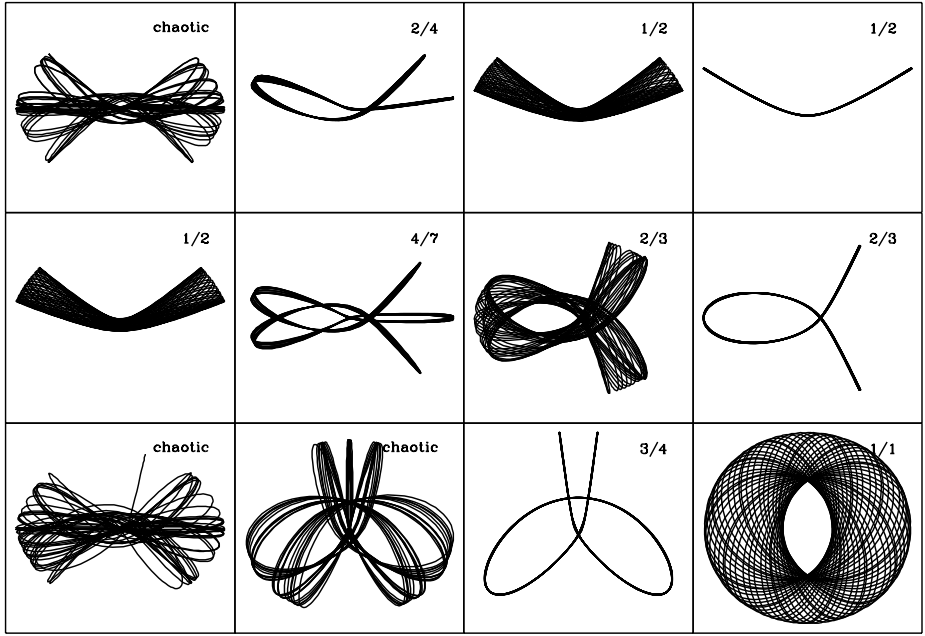

FIGURE 3. A set of representative orbits in a scale-free elliptic disk, with surface density distribution $\Sigma \propto 1/(x^2 + y^2/p^2)^{1/2}$, and $p = 0.6$. The long axis of the model is horizontal. All but the bottom-right (tube) orbit were started at zero velocity. The $x:y$ resonance that traps each orbit (if there is one) is indicated (Kuijken 1993).

Kuijken (1993) constructed scale-free elliptic disks with surface density $\Sigma \propto 1/(x^n + y^n/p^n)^{1/n}$, and $n = 2, 4$. This is the simplest selfconsistent problem that has the basic complexity of triaxial models with central cusps. All flat tube orbits are regular, but there are no regular flat boxes as in the elliptic disks with finite density core (Teuben 1986; de Zeeuw, Hunter & Schwarzschild 1987; Levine & Sparke 1994). The box orbits are replaced by flat regular boxlets and irregular (stochastic) orbits (Figure 3). Kuijken found that strongly elongated disks can not be constructed with the available orbits. Moderately elongated disks can be built, but the precise range of elongations where solutions exist depends somewhat on the angular resolution of the numerical grid on which the orbital densities are calculated, and also depends on the value of $n$ (see fig. 3 in Kuijken 1993).

Schwarzschild (1993) carried out a similar study of triaxial scale-free models with a $1/r^2$ density distribution, and a nearly ellipsoidal shape with axis ratios $q \leq p \leq 1$. He investigated six shapes, and used 600 orbits on a grid of modest angular resolution (48 cells in one octant). This work showed that models as flat as $q \sim 0.5$ can be built by populating only the regular orbits (three tube-orbit families, and the regular boxlets).

Flatter models, with $q \sim 0.3$, require inclusion of irregular orbits. These results are in harmony with Kuijken's experiments.

### 3.3. *Power-law models*

It has recently become clear that there is a third set of special triaxial models with interesting and useful properties. These are the so-called *power-law models*, with potentials stratified on similar ellipsoids with axis ratios $p$ and $q$, and a radial profile which is a power $\gamma - 2$ of $(R_c^2 + x^2 + y^2/p^2 + z^2/q^2)$, where $R_c$ is a core radius. When $\gamma = 2$ the potential is logarithmic, and is identical to that of the model studied by Binney (1981). A fifth parameter is the tumbling rate of the model, $\Omega_b$. The power-law models with $R_c \neq 0$ have cores, while those with $R_c = 0$ are scale-free. They allow efficient investigation of the observables of models with cores as well as cusps, as a function of the model parameters, and of the two viewing angles that define the direction of observation.

These models are the triaxial generalizations of the axisymmetric power-law models introduced and studied by Evans (1993, 1994). The latter have extremely simple DFs $f(E, L_z)$, with observable properties that can be calculated explicitly (Evans & de Zeeuw 1994). In addition, a multi-parameter set of anisotropic solutions of the Jeans equations exists, and the associated observables can be computed explicitly. De Zeeuw, Evans & Schwarzschild (1995) have identified the solutions that are most likely to correspond to physical DFs $f(E, L_z, I_3)$ for the axisymmetric scale-free logarithmic limit ($R_c = 0, \gamma = 2$). Work on the general case $\gamma \neq 2$ is in progress.

Multi-parameter Jeans solutions have been derived also for the triaxial models, and again their line-of-sight velocity dispersions can be written down explicitly (Carollo, de Zeeuw & Evans 1995). More work is needed to determine which of these Jeans solutions correspond to physical DFs.

### 3.4. *Numerical models with cusps and black holes*

Further insight in the dynamics of general triaxial systems has come from numerical construction attempts. Models built with Schwarzschild's method often use orbital densities derived by numerically integrating the equations of motion for 50–100 dynamical times. Over such timescales, irregular orbits behave very much like fuzzy regular orbits. On longer timescales, the density distributions of irregular orbits evolve. Eventually, they fill the entire part of the equipotential volume defined by the orbital energy not occupied by the regular orbits. Therefore, a dynamical model with stars on irregular orbits that have been integrated for short times is not strictly in equilibrium (cf. Binney 1982): it evolves on a timescale which is long compared to the dynamical time, and short compared to the two-body relaxation time.

The secular evolution driven by the irregular orbits is probably unimportant for the structure of galactic halos. The outer regions of galaxies are less than 50 dynamical times old. Furthermore, the potential is not likely to be very smooth, since it is perturbed by sporadic infall of satellite galaxies. The situation is different in the central cusps, where the dynamical time becomes very short. Since equipotential surfaces are generally less elongated than the associated density distribution, a triaxial model with many stars on irregular orbits will tend to evolve towards a more nearly axisymmetric shape on a time scale that is comparable to or shorter than a Hubble time (Gerhard 1986, 1987). All models that include orbital density contributions by irregular orbits that have been integrated for short times should therefore be treated with caution.

The worry about the use of irregular orbits is not restricted to purely scale-free models. Boxlets occur in many models with a central density cusp, or with a massive central black hole (Gerhard & Binney 1985; Udry & Pfenniger 1988; de Zeeuw & Pfenniger 1988; Binney & Petit 1989; Pfenniger & de Zeeuw 1989; Lees & Schwarzschild 1992). Different approaches have been followed in order to deal with the problem of irregular orbits. Some studies have simply ignored them in the construction of selfconsistent DFs, even though it is not easy to imagine a galaxy formation process that would do so. More recent investigations have used the fact that all the irregular orbits at one energy are really the same orbit, when integrated over sufficiently many dynamical periods. Merritt & Fridman (1995) suggested that (i) in the inner regions of a model, the orbit catalog needed for application of Schwarzschild's method should include only one irregular orbit at each energy, integrated for long times, and (ii) at radii beyond which the dynamical time is larger than about 1/50th of a Hubble time, one should include shorter integrations of pieces of the same irregular orbit. These authors applied these principles to two models with the same strongly triaxial shape ($q = 0.5, p = 0.8$), the first with a shallow cusp ($\gamma = 1$), and the second with a steep cusp ($\gamma = 2$). They found that only the model with the shallow slope can be succesfully reproduced with this way of treating the irregular orbits. The steep-cusped model can be constructed by using the regular orbits and only short integrations of the irregular orbits, but it evolves on longer timescales.

Although the number of numerical constructions of scale-free and realistic triaxial models is still limited, the results obtained so far suggest that models with cores or shallow central cusps may well exist for all relevant triaxial shapes, whereas those with steep cusps either evolve relatively fast, or are possible only for near-axisymmetric shapes.

### 3.5. *Tumbling triaxial systems*

Triaxial elliptical galaxies may well have slowly tumbling figures. The associated Coriolis force, which distinguishes between direct and retrograde motion "significantly complicates the situation" (Lynden–Bell 1962c). Other than the homogeneous ellipsoids (Freeman 1966a,b,c; Hunter 1974), no rotating equivalents are known of the separable models discussed in §3.1. The one exception (Vandervoort 1979) has an unphysical density distribution (Contopoulos & Vandervoort 1992). Nearly all our knowledge of the orbital structure in rotating systems is therefore based on numerical orbit calculations. Three-dimensional studies have concentrated mostly on tumbling around the short axis (see reviews by Binney 1987; Sellwood & Wilkinson 1993; de Zeeuw 1994). Most orbits are regular for slow tumbling rates, but minor orbit families and irregular orbits always appear. The directional non-uniqueness of stationary triaxial systems disappears in models with a rotating figure. The clockwise and counterclockwise branch of an orbit have different shapes. The long-axis tubes tip out of the plane containing the intermediate and short axis, with the two branches tipping in opposite directions, so that they have non-zero angular momentum about both the long and the short axis (Schwarzschild 1982; Vietri & Schwarzschild 1983). In order to reproduce a symmetric triaxial shape, the two branches of the long-axis tubes must be populated in equal numbers, so that the total angular momentum (tumbling + streaming) points along the intrinsic short axis. Conversely, net streaming around the long axis of a tumbling shape results in asymmetries (intrinsic twists). It follows that one cannot first reconstruct the density of a tumbling triaxial model, and then choose the mean streaming motions, as is possible for stationary shapes; both the even and the odd part of the DF are constrained by the density.

Very little work has been done on the construction of tumbling triaxial galaxy models. Schwarzschild's method has produced some rotating models with cores (Schwarzschild 1982; Pfenniger 1984; Vietri 1986; Zhao 1995), as have N-body simulations. The presence of a central density cusp or black hole in a tumbling triaxial system is expected to influence the orbital structure less strongly than in a stationary triaxial model, because the Coriolis force causes box orbits to acquire net angular momentum and to become centrophobic. Even so, a sufficiently large central mass concentration does dissolve triaxial bar models (Hasan, Pfenniger & Norman 1993; Norman, Sellwood & Hasan 1995). It is unknown whether a (fast)- tumbling triaxial system can support a steep cusp.

## 4. Implications

The main result of the theoretical work described in the previous section is that the orbital structure in a triaxial system depends on its central mass concentration. Thus, the internal dynamical structure differs in galaxies with shallow and steep cusps.

Recent observations with the Hubble Space Telescope (HST) have shown that the density profiles of all ellipticals approach a power-law form $\rho(r) \propto r^{-\gamma}$ at small radii $r$ (Crane et al. 1993; Ferrarese et al. 1994; Forbes et al. 1995; Lauer et al. 1995; Kormendy et al. 1995; Carollo et al. 1995). However, giant ellipticals have shallow cusps, while low-luminosity ellipticals have steep cusps (Kormendy et al. 1994). When combined with the theoretical work on model building, these observations are consistent with the knowledge that giant ellipticals are triaxial (Binney 1976), and suggest that low-luminosity ellipticals are likely to have near-axisymmetric shapes. There is one caveat to the latter statement: it is not yet known whether tumbling triaxial systems with steep cusps can be constructed. If this is possible, then the low-luminosity ellipticals may be triaxial. In this connection, it is interesting to investigate barred galaxies. These are definitely triaxial, and have substantial tumbling rates. Their luminosities are comparable to those of the smaller ellipticals. High spatial resolution studies of their central luminosity distributions are being collected to investigate the relation between triaxiality, tumbling rate, and central mass concentration.

It is not known whether the differences between the giant and small ellipticals were imposed by the galaxy formation process, or are the result of subsequent dynamical evolution, or both. Open questions include: (i) Was the role of dissipation more important in low-luminosity ellipticals than in the giants, so that they developed steeper cusps? (ii) Did mergers destroy pre-existing steep cusps, and formed triaxial systems with shallow cusps? (iii) Do the massive black holes that are known to be present in giant ellipticals with powerful radio sources result in a secular evolution towards axisymmetry, and if so, on what timescale? (iv) Do giant ellipticals have binary black holes, and do they cause the shallow cusps?

## 5. Future work

Progress in this field relies heavily on a combination of theoretical advances in dynamical modeling and observational improvements. There are a number of areas where progress can be made.

### 5.1. *Further exploration on the special families of triaxial models*

Despite the uncertainty about the physical validity of the solutions of the Jeans equations (§2), the difficulty of constructing distribution functions for triaxial galaxies leaves room for further investigation of the lowest velocity moments of their DFs. Very little work has been done in this area, even though, e.g., the Jeans equations for the separable models form a closed set, and were written down 35 years ago in Lynden-Bell's (1960b) PhD Thesis. Jeans solutions for the three special families of triaxial models discussed in §§3.1–3.3 will provide essential information on the variety of dispersion profiles expected for selfconsistent and non-selfconsistent models of different shapes and density profiles.

Jeans solutions must be complemented by further construction of DFs. For the separable models this can be done by Dejonghe's (1989) quadratic programming method, or variants thereof. It is also possible to use the iterative scheme developed by Robijn & de Zeeuw (1995) to compute three-integral DFs for separable models starting from the explicitly known maximum streaming DFs (Hunter & de Zeeuw 1992). These methods provide smooth DFs, and allow analysis of VPs for triaxial models with cores, as a function of position on the sky, and of the two axial ratios as well as the direction of observation.

### 5.2. *Construction of general triaxial models*

Schwarzschild's method allows a systematic exploration of the existence of triaxial equilibria as a function of cusp slope, of the axis ratios, of the tumbling rate, and of the presence and mass of a central black hole. The method can be extended to include as constraints not only the observed velocity dispersions (Richstone & Tremaine 1984, 1988), but also the entire VP shapes (Rix, de Zeeuw & Carollo 1995). In principle, construction of a tumbling triaxial model is just as easy as building a spherical model. In practice, the former requires a substantially larger investment of cpu time, because a much larger ($\sim 10^4$) number of orbits is needed to sample phase-space. Coverage of all viewing directions requires additional cpu. Even so, the speed and storage capabilities of current computers makes it possible to carry out a systematic investigation of these models. This large undertaking will benefit from the insight provided by the analysis of the special families of triaxial models.

### 5.3. *N-body simulations of triaxial galaxies*

An accurate treatment of secular evolution, such as driven by the growth of a central black hole, can be investigated with current N-body codes. Issues to be addressed include the role of the central black hole in determining the cusp slope and shape of the host galaxy, and the effect of the capture of another galaxy, possibly with its own black hole.

The central stellar density in some ellipticals with cusps is so high that the two-body relaxation time for stellar encounters becomes significantly less than a Hubble time, so that the collisionless approximation fails sufficiently close to the center. This has received little attention in the context of galactic nuclei, although much work has been done on the evolution of globular clusters with cusped density distributions (e.g., Hut et al. 1992). This is a non-trivial problem, that needs to be solved urgently.

Little is known about the importance of dynamical instabilities in limiting the possible triaxial equilibrium models available to galaxies. In principle, linear stability analysis can be carried out for the separable triaxial models, but this is currently somewhat impractical for technical reasons. N-body simulations are essential in this area.

### 5.4. *Spectroscopic observations*

Optical imaging studies of elliptical galaxies have matured in the last decade, because of the large body of ground-based work carried out with CCD's (e.g., Kormendy & Djorgovski 1989), and because of the high spatial resolution that can be achieved with HST. By contrast, spectroscopic studies are still limited. They have relied largely on long-slit spectrographs. In one exposure, these provide the line-of-sight velocities and line-strengths along a one-dimensional cut across the galaxy image. Collecting the spectral information along more than a few position angles in this way is expensive in telescope time. On the other hand, the realization that many galaxy components are triaxial in shape, and that therefore their velocity fields and line-strength distributions can display a rich structure, makes *two-dimensional* (integral field) spectroscopy essential for deriving the dynamical structure of these systems, and for understanding their formation.

Three complementary spectroscopic investigations are needed:

• Ground-based long-slit kinematic studies of nearby low-luminosity ellipticals, and of more distant giants. The statistics of minor axis rotation can be used, in combination with accurate measurement of the distribution of ellipticities, to constrain the intrinsic shapes of these systems. This is achievable with existing instrumentation.

• Highest spatial resolution spectroscopy in the bright nuclear regions. This allows a study of the stellar kinematical signature of massive central black holes, especially in quiescent, i.e., non-active galaxies. A substantial instrumental effort is under way to make this line of research possible. The next upgrade of HST will provide long-slit spectra with 0.1 arcsec resolution. Integral field spectrographs with small fields of view but high spatial resolution (0.1–0.25 arcsec, by use of adaptive optics) are currently being constructed at a number of major observatories, including the Canada-France-Hawaii Telescope, and the William Herschel Telescope.

• Intermediate spatial resolution spectroscopy over the entire optical extent of elliptical galaxies. This will provide two-dimensional kinematics, which will significantly constrain the internal orbital structure, as well as maps of line-strengths. Much progress can be made by combination of the stellar motions with the physical properties of the stars, such as age and metallicity. The first step is to extend Schwarzschild's construction technique to incorporate also line-strength measurements. This means that the model needs to be constructed from orbits that not only cover phase-space, but also come in different flavors, i.e., different values of the line-strength. The extra constraints provided by the (two-dimensional) line-strength observations will then allow a physical determination of the intrinsic line-strength distribution in the galaxy. This will be a strong constraint on formation scenarios.

No integral field spectrograph exists with the large field of view required for this aim, even though it can be built fairly easily. Plans to do so are under way.

## 6. Concluding remark

Most galaxy components are very likely triaxial. Their internal dynamical structure is rich, and holds important information on their formation. Methods to deduce this internal structure now exist, and need to be applied to delineate the full variety of stable equilibrium models that can be constructed. The next few years will see much progress in this area, based on combination of these theoretical developments with state-of-the-art imaging and spectroscopy.

It is a great pleasure to thank Donald Lynden-Bell for the role he has played in my career. Marcella Carollo enthusiastically helped with the preparation of this review. Roeland van der Marel and Martin Schwarzschild commented on the manuscript, and Richard Arnold kindly provided some of the figures. This paper was written at the Institute for Advanced Study, with support from NSF Grant PHY 92-45317.

REFERENCES

AARSETH, S. & BINNEY, J.J. 1976 *MNRAS* **185**, 227–243.
AGUILAR, L.A. 1988 *Celest. Mech.* **41** 3–37.
ARNOLD, R.A., DE ZEEUW, P.T. & HUNTER C. 1994 *MNRAS* **271**, 924–948.
BARNES, J.E. 1992 *ApJ* **393**, 484–507.

BARNES, J.E. & HERNQUIST, L. 1992 *ARA&A* **30**, 705–742.
BERTIN, G. & STIAVELLI, M. 1993 *Rep. Progr. Phys.* **56**, 493–556.
BINNEY, J.J. 1976 *MNRAS* **177**, 19–29.
BINNEY, J.J. 1981 *MNRAS* **196**, 455–467.
BINNEY, J.J. 1982 *MNRAS* **201**, 15–19.
BINNEY, J.J. 1987 *IAU Symposium No. 127, Structure and Dynamics of Elliptical Galaxies*, ed. P.T. de Zeeuw (Dordrecht: Reidel), p. 229–239.
BINNEY, J.J. & TREMAINE S.D. 1987 *Galactic Dynamics* (Princeton University Press).
BINNEY, J.J. & PETIT, J.-M. 1989 *Dynamics of Dense Stellar Systems*, ed. D.R. Merritt (Cambridge Univ. Press), p. 43–55.
BLITZ, L. & SPERGEL, D.N. 1991 *ApJ* **379**, 631–638.
CAROLLO, C.M., FRANX, M., FORBES, D.A. & ILLINGWORTH, G.D. 1995 *AJ* submitted.
CAROLLO, C.M., DE ZEEUW, P.T. & EVANS, N.W. 1995 preprint.
CONTOPOULOS, G. 1960 *Z. Astroph.* **49**, 273–291.
CONTOPOULOS, G. & VANDERVOORT, P.O. 1992 *ApJ* **389**, 118–128.
CRANE, P., STIAVELLI, M., KING, I.R., DEHARVENG, J.M., ALBRECHT, R., BARBIERI, C., BLADES, J.C., BOKSENBERG, A., DISNEY, M.J., JAKOBSEN, P., KAMPERMAN, T.M., MACCHETTO, F., MACKAY, C.D., PARESCE, F., WEIGELT, G., BAXTER, D., GREENFIELD, P., JEDRZEJEWSKI, R., NOTA., A. & SPARKS, W.B. 1993 *AJ* **106**, 1371–1393.
DEHNEN, W. & GERHARD, O.E. 1994 *MNRAS*, 268, 1019–1032.
DEJONGHE, H.B. 1986 *Phys. Rep.* **133**, 217–313.
DEJONGHE, H.B. 1989 *ApJ*, **343**, 113–124.
DEJONGHE, H.B. 1992 *ESO/EIPC Workshop on Structure, Dynamics and Chemical Evolution of Elliptical Galaxies*, eds I.J. Danziger, W.W. Zeilinger & K. Kjär (Garching: ESO), 337–346.
DEJONGHE, H.B. & DE ZEEUW, P.T. 1988 *ApJ*, **333**, 90–129.
DEJONGHE, H.B. & LAURENT, D. 1991 *MNRAS*, **252**, 606–636.
DEJONGHE, H.B., DE BRUYNE, V., VAUTERIN, P. & ZEILINGER, W.W. 1995 *A&A* in press.
DE ZEEUW, P.T. 1985a *MNRAS* **215**, 731–760.
DE ZEEUW, P.T. 1985b *MNRAS* **216**, 273–334.
DE ZEEUW, P.T. 1985c *MNRAS* **216**, 599–612.
DE ZEEUW, P.T. 1994 *The Formation of Galaxies*, eds C. Muñoz–Tuñon & F. Sanchez (Cambridge University Press), 231–316.
DE ZEEUW, P.T., HUNTER, C. & SCHWARZSCHILD, M. 1987 *ApJ* **317**, 607–636.
DE ZEEUW, P.T. & LYNDEN–BELL, D. 1985 *MNRAS* **215**, 713–730.
DE ZEEUW, P.T., PELETIER, R.F. & FRANX, M. 1986 *MNRAS* **221**, 1001–1022.
DE ZEEUW, P.T. & PFENNIGER, D. 1988 *MNRAS* **235**, 949–996.
DE ZEEUW, P.T. & FRANX, M. 1991 *ARAA* **29**, 239–274.
DE ZEEUW, P.T., EVANS, N.W. & SCHWARZSCHILD, M. 1995 *MNRAS* submitted.
EDDINGTON, A.S. 1915 *MNRAS* **76**, 37–60.
EVANS, N.W. 1993 *MNRAS* **260**, 191–201.
EVANS, N.W. 1994 *MNRAS* **267**, 333–360.
EVANS, N.W. & DE ZEEUW, P.T. 1994 *MNRAS* **271**, 202–221.
FERRARESE, L., VAN DEN BOSCH, F.C., JAFFE, W., FORD, H.C. & O'CONNELL, R.W. 1994. *AJ* **108**, 1598–1609.
FORBES, D.A., FRANX, M. & ILLINGWORTH, G.D. 1995 *AJ* 109, 1988–2002.
FRANX, M. 1988 *MNRAS* **231**, 285–308.
FRANX, M. & ILLINGWORTH, G.D. 1988 *ApJL* **327**, 55–60.

FRANX, M., ILLINGWORTH, G.D.& DE ZEEUW, P.T. 1991. *ApJ* **383**, 112–134.
FREEMAN, K.C. 1966a *MNRAS* **133**, 47–62.
FREEMAN, K.C. 1966b *MNRAS* **134**, 1–14.
FREEMAN, K.C. 1966c *MNRAS* **134**, 15–23.
FRENK, C.S., WHITE, S.D.M., DAVIS, M. & EFSTATHIOU, G.P. 1988 *ApJ*, **327**, 507–525.
GERHARD, O.E. 1985 *A&A* **151**, 279–296.
GERHARD, O.E. 1986 *MNRAS* **219**, 373–386.
GERHARD, O.E. 1987 *IAU Symposium 127: Structure and Dynamics of Elliptical Galaxies*, ed. P.T. de Zeeuw (Dordrecht: Reidel), p. 241–248.
GERHARD, O.E. 1991 *MNRAS* **250**, 812–830.
GERHARD, O.E. & BINNEY, J.J. 1985 *MNRAS* **216**, 467–502.
HASAN, H., PFENNIGER, D. & NORMAN, C.A. 1993 *ApJ* **409**, 91–109.
HUNTER, C. 1974 *MNRAS* **166**, 633–648.
HUNTER, C. & DE ZEEUW, P.T. 1992 *ApJ* **389**, 79–117.
HUNTER, C. & QIAN, E.E. 1993 *MNRAS* **262**, 401–428.
HUT, P., MCMILLAN, S., GOODMAN, J., MATEO, M., PHINNEY, E.S., PRYOR, T., RICHER, H., VERBUNT, F. & WEINBERG, M. 1992 *PASP* **104**, 981-1034.
JEDRZEJEWSKI, R.I. & SCHECHTER, P.L. 1988 *ApJL* **330**, 87–91.
KORMENDY, J. 1982 *Morphology and Dynamics of Galaxies*, eds L. Martinet & M. Mayor (Sauverny: Geneva Observatory), 115–288.
KORMENDY, J., & DJORGOVSKI, S. 1989 *ARAA* **27**, 235–277.
KORMENDY, J., DRESSLER, A., BYUN, Y.I., FABER, S.M., GRILLMAIR, C., LAUER, T.R., RICHSTONE, D.O. & TREMAINE, S.D. 1994 *Dwarf Galaxies*, eds G. Meylan & P. Prugniel (Garching, ESO), p. 147.
KORMENDY, J., BYUN Y.I., AJHAR, E.A., LAUER, T.R., DRESSLER, A., FABER, S.M., GRILLMAIR, C., GEBHARDT, K., RICHSTONE, D.O. & TREMAINE, S.D. 1995 *IAU Symposium 171: New Light on Galaxy Evolution*, eds R. Bender & R.L. Davies (Dordrecht: Kluwer), in press.
KUIJKEN, K. 1993 *ApJ* **409**, 68–74.
KUZMIN, G.G. 1956 *Astr. Zh.* **33**, 27–45.
KUZMIN, G.G. 1973 *Dynamics of Galaxies and Clusters*, ed. T.B. Omarov (Alma Ata: Akad. Nauk. Kaz. SSR), p. 71–75 (transl. in IAU Symposium 127, *Structure and Dynamics of Elliptical Galaxies*, ed. P.T. de Zeeuw (Dordrecht: Reidel), p. 553–557).
LAUER, T.R., AJHAR E.A., BYUN, Y.I., DRESSLER, A., FABER, S.M., GRILLMAIR, C., KORMENDY, J., RICHSTONE, D.O. & TREMAINE, S. 1995 *AJ*, in press.
LEES, J.F. & SCHWARZSCHILD, M. 1992 *ApJ* **384**, 491–501.
LEVINE, S. & SPARKE, L.S. 1994 *ApJ* **428**, 493–510.
LEVISON, H.F. & RICHSTONE, D.O. 1985 *ApJ* **295**, 340–348.
LYNDEN–BELL, D. 1960a *MNRAS* **120**, 204–213.
LYNDEN–BELL, D. 1960b *Stellar and Galactic Dynamics*, PhD Thesis, Cambr. Univ.
LYNDEN–BELL, D. 1962a *MNRAS* **123**, 447–458.
LYNDEN–BELL, D. 1962b *MNRAS* **124**, 1–9.
LYNDEN–BELL, D. 1962c *MNRAS* **124**, 95–123.
MERRITT, D.R. & FRIDMAN, T. 1995 preprint.
MIRALDA-ESCUDÉ, J. & SCHWARZSCHILD M. 1989 *ApJ* **339**, 752–762.
NORMAN, C.A., SELLWOOD, J.A. & HASAN, H. 1995 preprint.
OLLONGREN, A. 1962 *BAN* **16**, 241–295.
PFENNIGER, D. 1984 *A&A* **141**, 171–188.

PFENNIGER, D. & DE ZEEUW, P.T. 1989 *Dynamics of Dense Stellar Systems*, ed. D.R. Merritt (Cambridge Univ. Press), p. 81–88.
QIAN, E.E., DE ZEEUW, P.T., VAN DER MAREL, R.P. & HUNTER, C. 1995 *MNRAS* **274**, 602–622.
RICHSTONE, D.O. 1980 *ApJ* **238**, 103–109.
RICHSTONE, D.O. 1984 *ApJ* **281**, 100–111.
RICHSTONE, D.O. & TREMAINE, S.D. 1984 *ApJ* **286**, 27–37.
RICHSTONE, D.O. & TREMAINE, S.D. 1988 *ApJ* **327**, 82–88.
RIX, H.-W., DE ZEEUW, P.T. & CAROLLO, C.M. 1995 preprint.
ROBIJN, F.H.A. & DE ZEEUW, P.T. 1995 *MNRAS* in press.
SCHWARZSCHILD, M. 1979 *ApJ* **232**, 236–247.
SCHWARZSCHILD, M. 1982 *ApJ* **263**, 599–610.
SCHWARZSCHILD, M. 1993 *ApJ* **409**, 563–577.
SELLWOOD, J., & WILKINSON, A. 1993 *Rep. Progr. Phys.* **56**, 173–255.
STÄCKEL, P. 1890 *Math. Ann.* **35**, 91–103.
STATLER, T.S. 1987 *ApJ* **321**, 113–152.
STATLER, T.S. 1991a *AJ* **102**, 882–892.
STATLER, T.S. 1991b *ApJ* **382**, 11–15.
STATLER, T.S. 1994a *ApJ* **425**, 458–480.
STATLER, T.S. 1994b *ApJ* **425**, 500–529.
TEUBEN, P.J. 1987 *MNRAS* **215**, 815–841.
TOOMRE, A. 1982 *ApJ* **259**, 535–543.
UDRY, S. & PFENNIGER, D. 1988 *A&A* **198**, 135–149.
VANDERVOORT, P.O. 1979 *ApJ* **232**, 91–105.
VIETRI, M. 1986 *ApJ* **306**, 48–63.
VIETRI, M. & SCHWARZSCHILD, M. 1983 *ApJ* **269**, 487–499.
WILKINSON, A. & JAMES, R.A. 1982 *MNRAS* **199**, 171–196.
ZHAO, H.S. 1995 *MNRAS* in press.